\documentclass[twocolumn,secnumarabic,amssymb, prl,superscriptaddress]{revtex4-2} 

\usepackage{lipsum}
\usepackage{amsfonts,amssymb,amsmath}
\usepackage{graphicx}
\usepackage{bm}
\usepackage{xcolor,calc}
\usepackage{siunitx}
\usepackage{nicefrac}

\definecolor{color_comment}{rgb}{0.3, 0.8, 0.3}

\definecolor{color_out}{rgb}{0.7, 0.7, 0.7}

\definecolor{color_new}{rgb}{0.8, 0.3, 0.3}

\begin{document}

	\author{Katharina Senkalla}
	\thanks{\href{mailto:katharina.senkalla@uni-ulm.de}{katharina.senkalla@uni-ulm.de}}
	\affiliation{Institute for Quantum Optics, Ulm University, Albert-Einstein-Allee 11, 89081 Ulm, Germany}
	\author{Genko Genov}
	\affiliation{Institute for Quantum Optics, Ulm University, Albert-Einstein-Allee 11, 89081 Ulm, Germany}
	\author{Mathias H. Metsch}
	\affiliation{Institute for Quantum Optics, Ulm University, Albert-Einstein-Allee 11, 89081 Ulm, Germany}
	\author{Petr Siyushev}
	\affiliation{Institute for Quantum Optics, Ulm University, Albert-Einstein-Allee 11, 89081 Ulm, Germany}	
	\affiliation{3rd Institute of Physics, Center for Applied Quantum Technologies University of Stuttgart, Stuttgart, Germany}
	\affiliation{Institute for Materials Research (IMO), Hasselt University, Wetenschapspark 1, B-3590 Diepenbeek, Belgium}
	\author{Fedor Jelezko}
	\affiliation{Institute for Quantum Optics, Ulm University, Albert-Einstein-Allee 11, 89081 Ulm, Germany}

	\title{Germanium Vacancy in Diamond Quantum Memory Exceeding 20 ms}
	
	\date{\today}
	
	\begin{abstract}
		Negatively charged group-IV defects in diamond show great potential as quantum network nodes due to their efficient spin-photon interface. However, reaching sufficiently long coherence times remains a challenge.
		In this work, we demonstrate coherent control of germanium vacancy center (GeV) at millikelvin temperatures and extend its coherence time by several orders of magnitude to more than 20 ms. We  model the magnetic and amplitude noise as an Ornstein-Uhlenbeck process, reproducing the experimental results well. The utilized method paves the way to optimized coherence times of group-IV defects in various experimental conditions and their successful applications in quantum technologies.
	\end{abstract}
	
	\maketitle

	Quantum networks have the potential to enhance the way we communicate and process information by enabling new technologies such as distributed quantum computing, enhanced sensing, and secure quantum communication \cite{kimble2008quantum,Childress2005Faulttolerant,Gottesman2012telescopes,komar2014quantum,Monroe2014quantumcomputer}.
	Specifically, long-distance quantum communication remains an open challenge as it requires qubits that act as a long-lived quantum memory for efficient entanglement distribution. \\
	Recent studies have reported the great potential of negatively charged group-IV defects in diamond as a quantum network node \cite{bradac2019quantum,Ruf2021QuantumNetworks}.
	The defects share outstanding optical properties such as high flux of coherent photons (Debye-Waller factor up to $\sim$70\%), Fourier-transform-limited optical transitions, and exceptional spectral stability imposed by the inversion symmetry of the defect's structure~\cite{bradac2019quantum}.
	Spectral stability is essential for the integration into nanophotonic devices, which has already been demonstrated for various defects~\cite{Sipahigil_S2016,Bhaskar_PRL2017,Rugar_ACSP2020}. 
	In order to satisfy all requirements for a network node, the systems should, moreover, provide access to a well-controllable spin qubit with a long quantum memory time.
	Such control has been demonstrated with silicon-vacancy centers (SiV) in diamond, with memory times approaching $\sim$10~ms~\cite{sukachev2017silicon}.
	Despite the showcased achievements, the SiV's electron spin suffers of phonon-mediated decoherence due to its small orbital ground state splitting (48~GHz), as shown in~\cite{Jahnke_NJP2015,pingault2017coherent}.
	To mitigate this effect, approaches such as strain engineering of the defects~\cite{Meesala_PRB2018,Stas2022Robust}, or operation in dilution refrigerators \cite{sukachev2017silicon,becker2018alloptical} have been explored.
	Strain may potentially impact the spectral stability of the defect and introduce additional complexity, so operating at low temperatures remains the preferred solution. 
	However, performing experiments in dilution refrigerators requires a careful adjustment of the induced heat load as the cooling power is limited. \\
	These challenges have motivated efforts for the investigation of other group-IV defects.
	These defects provide not only enhanced optical properties, such as higher coherent flux of photons, but also an increasing spin-orbit splitting across the group, which allows operation at elevated temperatures \cite{Ruf2021QuantumNetworks,bradac2019quantum}.
	The germanium vacancy (GeV) is considered as a promising alternative.
	The fabrication is relatively easy \cite{siyushev2017gev,Bhaskar_PRL2017}, similarly to SiV, preserving good optical characteristics.
	However, the suppression of phonon relaxation at a few hundred millikelvin is more than four orders of magnitude higher compared to SiV.
	This enables the use of strong microwave (MW) fields for coherent control.\\
	In this Letter, we demonstrate for the first time efficient initialization, readout and coherent control of a negatively charged GeV center at temperatures below 300~mK. 
	At these temperatures, the phonon relaxation process is suppressed and we observe spin noise limited coherence time of the order of $T_2^{\ast}\approx\,1.43\,\mu$s. 
	We prolong the quantum memory time by several orders of magnitude  to more than $ 20\,\text{ms}$ by dynamical decoupling (DD) protocols.
	The achieved memory time exceeds the one of SiV by a factor of two \cite{sukachev2017silicon}, demonstrating that GeV is a viable alternative for quantum memory applications. 
	Our analysis shows that magnetic noise due to interactions with the spin environment and power fluctuations of the driving field can account for the most of the observed decoherence. 
	The noise is modeled as an Ornstein-Uhlenbeck process \cite{GillespieAJP1996} resulting in good agreement of the simulations and the experimental results.
	These findings allow for the design of efficient control strategies for extending the coherence times even further, e.g. by using higher-order DD sequences and tailoring the interpulse time separation \cite{ajoy2011optimal,Genov2017PRL,ezzell2023dynamical}. 
	The demonstration of efficient initialization, readout and coherent control in combination with long memory times of negatively charged GeV centers opens the door for multiple quantum technology applications, e.g., in quantum communication and quantum information. 
	
	\textit{Experimental setup and results.---}
	%
	We perform the experiments on a $\langle1,1,1\rangle$-oriented synthetic diamond grown via high-pressure high-temperature method with Ge incorporation during this process \cite{palyanov2015germanium}.
	During this high-pressure high-temperature growth process germanium (Ge) is naturally incorporated into the diamond, leading to the formation of GeV without requiring any additional treatments. 
	To optimize the collection efficiency we fabricate a solid immersion lens with 10\,$\mu$m diameter into the diamond, and position a 20-$\mu$m-thick wire nearby which delivers the microwave field.
	The sample is mounted on a cold finger of an optical dilution refrigerator combined with a home-built confocal microscope for individual addressing of GeV centers.
	The superconducting vector magnet allows for arbitrary alignment of a magnetic field with respect to the principal axis of the defect.
	Further details about the device preparation can be found in \cite{SM}.
	\begin{figure}[t!]
		\includegraphics[width=\columnwidth]{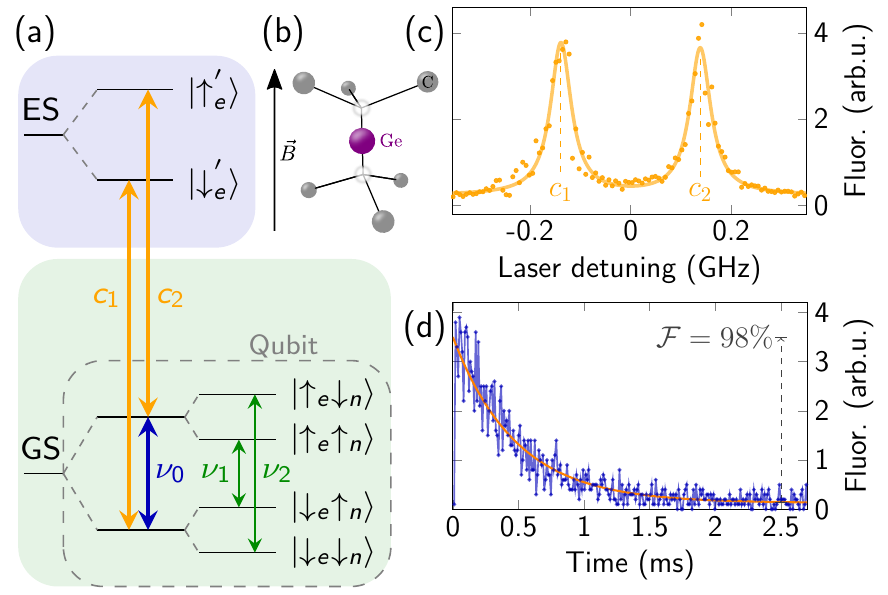}
		\caption{
			(a) Reduced energy level scheme showing the lower orbital branches in the ground (green) and excited (blue) state manifold.
			The electron spin becomes accessible by applying an external magnetic field $B=100\,\text{mT}$ (b) through the optical transitions $c_{1,2}$.
			(c) Corresponding PLE spectrum under constant MW repumping at frequency $\nu_0$.
			Slight misalignment of $\Vec{B}$ to the GeV axis allows for efficient optical initialization of 98\% within 1\,ms (d).
		}
		\label{Fig:Physical_system}
	\end{figure}
	\\Figure~\ref{Fig:Physical_system}(a) shows a reduced energy level diagram of the GeV center, emphasizing the relevant sublevels for the spin dynamics.
	At temperatures $T<\frac{h\Delta_{g}}{k_B}$, with $h$ as Planck's constant, $\Delta_{g}$ as ground state splitting, and $k_B$ as the Boltzmann constant, the orbital relaxation process becomes exponentially suppressed.
	We maintain a temperature below $300\,\text{mK}$ in all experiments \cite{SM}, so we consider only the lower orbital branches of the ground state (GS) and excited state (ES) manifold.
	To access the spin degree of freedom we apply a magnetic field $B=100\,\text{mT}$ and exploit the difference of the Zeeman splitting in GS and ES for resonant optical addressing.
	Figure~\ref{Fig:Physical_system}(c) shows the photoluminescence excitation (PLE) spectrum of the optical transitions $c_{1,2}$ in Fig.~\ref{Fig:Physical_system}(a) using an optical power of 2\,nW directed into the cryostat.
	Their spin-conserving nature leads to long cyclicity and, thus, to a low spin polarization rate.
	We choose a slightly misaligned magnetic field to induce spin state mixing, which reduces the required optical pumping time to 1\,ms with 98\% initialization fidelity.
	Figure~\ref{Fig:Physical_system}(d) shows the corresponding time-dependent luminescence trace using transition $c_2$.
	The fully initialized spin cannot be further driven by a field with frequency $c_2$ and, thus, is referred to as the ``dark state".
	As in millikelvin environment relaxation processes of the electron spin do not occur on relevant timescales \cite{sukachev2017silicon,becker2018alloptical}, a repumping scheme is required to resolve both transitions in PLE measurements.
	This can be achieved either by using an additional pump laser~\cite{becker2018alloptical} or by resonantly flipping the spin using microwave control.\\
	We determine the resonance frequency $\nu_0$ by sweeping a microwave around 3\,GHz after initialization in the dark state.	
	When the MW frequency matches the Zeeman splitting between the ground states, the population and, thus, the fluorescence are restored leading to an optically detectable magnetic resonance (ODMR).
	We note that, within one orbital branch, the orbital states are orthogonal which would, in principle, prevent direct microwave driving. 
	However, the GeV center under investigation shows a signature of strain with $\Delta_g=181$\,GHz \cite{SM}, so the orbital states mix and the transitions become allowed.
	We refer to Figure S.6 in \cite{SM} for an extended level scheme and corresponding PLE measurements.
	%
	\begin{figure}[t!]
		\hspace*{-0.4cm}\includegraphics[width=1.1\columnwidth]{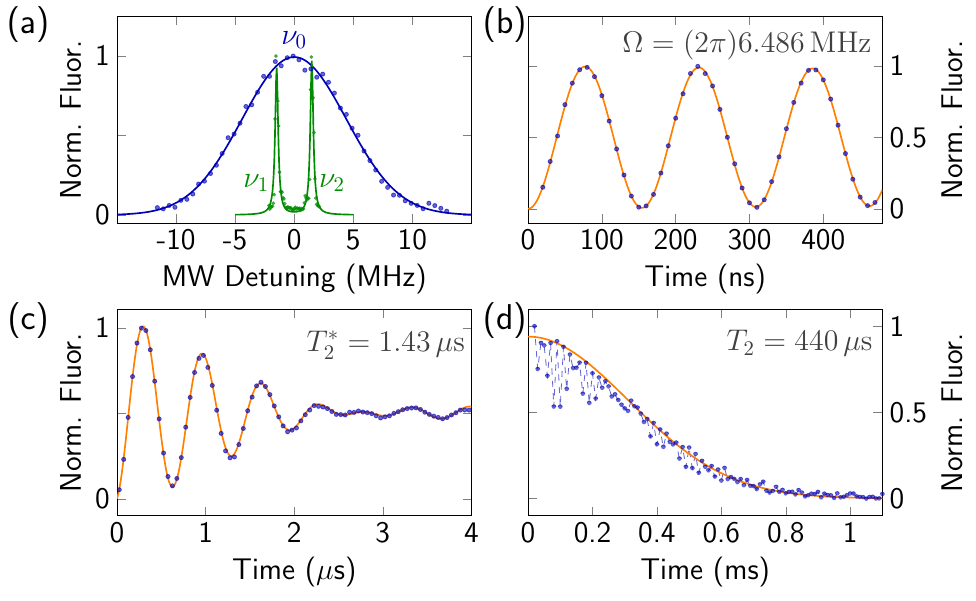}
		\caption{
			(a) ODMR spectrum (shown in green) shows transition frequencies $\nu_{1,2}$ with $300\,\text{kHz}$ linewidth. 
			The separation by $ 2.98\,\text{MHz}$ indicates a strongly coupled $^{13}$C.
			For further measurement, the driving field $\nu_0=3.066\,\mathrm{GHz}$ was chosen such that both transitions are equally covered (shown in blue).
			(b) Corresponding Rabi oscillations with a frequency of $\Omega = (2\pi) 6.486\,\mathrm{MHz} $.
			(c) Ramsey interference measurement reveals $T_2^*=1.43\,\mu s$. 
			(d) Hahn echo decay measurement yields spin-noise-limited $T_2=440\,\mu\text{s}$.
		}
		\label{Fig:Spin_properties}
	\end{figure}
	%
	\\The ODMR results shown in Fig. \ref{Fig:Spin_properties}(a) are conducted in a pulsed manner \cite{Walsworth2020RMP}.
	We observe a splitting of $2.98\,\text{MHz}$ due to hyperfine coupling to a nearby $^{13}$C nuclear spin with the linewidths of $\nu_1$ and $\nu_2$ approximately $300\,\text{kHz}$ [green curve in Fig.~\ref{Fig:Spin_properties}(a)].
	We set the frequency of the driving field to $\nu_0=3.066\,\text{GHz}$ for the further measurements, so it covers equally well both transitions $\nu_{1,2}$ due to power broadening [Fig. \ref{Fig:Spin_properties}(a) blue curve].
	We observe Rabi oscillations in Fig. \ref{Fig:Spin_properties}(b) and estimate a Rabi frequency of $\Omega = (2\pi)6.486\,\text{MHz}$ at 36\,dBm input power into the cryostat, inferring a $\pi$ pulse duration of 77.09\,ns.
	At the start of each experiment, the system is initialized in the dark state and subsequently coherently controlled using rectangular $\pi$ and $\frac{\pi}{2}$ pulses with durations determined from the Rabi measurement.
	\\
	We investigate the electron coherence time utilizing Ramsey interferometry, consisting of two $\frac{\pi}{2}$ pulses and a variable interpulse delay.
	This and all following measurements are performed in an alternating manner where we change the phase of the latter $\frac{\pi}{2}$ pulse between $X$ ($0^\circ$) and $-X$ ($180^\circ$) to project onto the dark and bright states. 
	We consider then the differential signal between them to reduce the effect of laser fluctuations and normalize to the maximum fluorescence difference unless otherwise stated \cite{SM}.
	By fitting the spin decay in Fig. \ref{Fig:Spin_properties}(c) we find the inhomogeneous spin dephasing time of $T_2^*\approx 1.43\,\mu\text{s}$.
	The oscillatory signal arises due to the microwave frequency detuning from the transitions $\nu_{1,2}$ to $\nu_0$, confirming the hyperfine coupling of $2.98\,\text{MHz}$. \\
	By operating in a temperature regime in which the phonon-induced transitions between orbital states are suppressed, the dephasing is mainly caused by magnetic noise. 
	For this regime, the coherence time $T_2$ can be significantly extended compared to $T_2^*$ using dynamical decoupling (DD) protocols. 
	These include $\pi$ pulses that periodically aim to refocus the phase accumulated by the GeV center due to interactions with the surrounding nuclear and electron spin bath \cite{Suter2016RevModPhys}.
	The Hahn echo is the simplest DD protocol having one additional $\pi$ pulse in the free evolution time between the two $\frac{\pi}{2}$ pulses of the Ramsey experiment. 
	Figure~\ref{Fig:Spin_properties}(d) shows the corresponding decay curve, which exhibits a modulation that can be attributed to the entangling and disentangling to the $^{13}$C spin bath.
	However, the modulation contrast is low due to the high Larmor precession frequency ($\approx1.03\,\text{MHz}$ \cite{SM}) and the undersampling of the pulse separation $\tau$. 
	From the fit we extract the spin coherence time $\text{T}_{2}\approx 440\,\mu\text{s}$.
	This timescale is consistent with the theoretically predicted hyperfine noise limit for diamonds with natural abundance of $^{13}$C~\cite{Hall_PRB2010}.\\
	The coherence time can be further extended using DD with multiple refocusing pulses  \cite{Viola1999PRL,Suter2016RevModPhys}.
	First, we apply the Carr-Purcell-Meiboom-Gill (CPMG) sequence consisting of an even number of $\pi$ pulses shifted by 90$^{\circ}$ with respect to the $\frac{\pi}{2}$ pulses [Fig. \ref{Fig:CPMG_tau}(a)] \cite{carr1954effects,MeiboomGill1960}. 
	In a first series of DD measurements, we keep the number of repetition pulses $N$ constant, while sweeping the interpulse delays $\tau$.
	Figure~\ref{Fig:CPMG_tau}(b) illustrates the extension of the coherence time with increasing $N$.
	The signal shows pronounced dips due to coupling to $^{13}$C, as exemplary shown for $N=2$ in the inset in Fig.~\ref{Fig:CPMG_tau}(b).
	Fitting the various datasets to a stretched exponential exhibits memory times up to $7$ times longer than for the Hahn echo. 
	%
	\begin{figure}[t!]
		\includegraphics[width=\columnwidth]{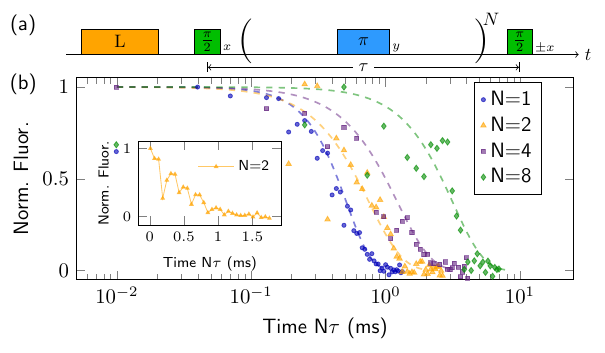}
		\caption{
			(a) CPMG sequence with $ N $ refocusing $ \pi $ pulses separated by $\tau$. 
			The phase of each pulse is indicated in a subscript.
			(b) Decay curves with fixed $N=\{1,2,4,8\}$ and increasing $\tau$. 
			Dashed lines show fits of the envelopes to $\exp{[-(N\tau/T_2)^{\beta}]}$ with $\beta$ a free parameter.
			Inset: Enlargement in of $N=2$ measurement, showing regular dips due to entangling and disentangling to a $^{13}$C.
		}
		\label{Fig:CPMG_tau}
	\end{figure}
	\\Experiments where the interpulse delay $\tau$ is varied and the number of pulses $N$ is kept constant are typically used to probe the spin environment noise spectrum \cite{Degen2017RMP}. 
	However, in quantum memory experiments we usually choose an optimal interpulse delay $\tau$ and vary the number of pulses $N$ \cite{ajoy2011optimal,SouzaPRL2011,Pascual-WinterPRB2012,Bar-GillNatComm2013,Zhong2015,Genov2017PRL,ezzell2023dynamical}. 
	This allows for memory time optimization and readout  at arbitrary times when the quantum state is refocused.
	To explore the limit of the spin memory time, we thus vary the order $N$ for the CPMG and $XY8$ sequences [Fig. \ref{Fig:CPMG_XY8_N}(a)], keeping constant a pulse spacing of $\tau=100\,\mu$s for which the Hahn echo decay is negligible. 
	We note that the pulse separation can be optimized further by tailoring it to the specific DD sequence \cite{ezzell2023dynamical} and by avoiding unwanted coupling to the $^{13}$C bath, e.g., due to spurious harmonics \cite{Loretz2015PRX}. 
	We choose the CPMG and $XY8$ sequences because the former is highly robust to errors when the initial $\pi/2$ pulse is shifted by $90^\circ$ with respect to the subsequent $\pi$ pulses. 
	However, its fidelity suffers if the initial $\pi/2$ pulse has the same phase \cite{carr1954effects,MeiboomGill1960,SouzaPRL2011,genov2018pra,SM}. 
	We thus also apply the widely used $XY8$ sequence, consisting of eight consecutive $\pi$ pulses, with  phases of $0^\circ$ ($X$) and $90^\circ$ ($Y$), as depicted in Fig. \ref{Fig:CPMG_XY8_N}(c). 
	It is robust to pulse errors and has a high fidelity for unknown initial states \cite{gullion1990new,SouzaPRL2011,genov2018pra}.
	A discussion on the fidelities of the pulses, CPMG and XY8 is included in \cite{SM}. 
	We note that even more advanced sequences can be applied like the Knill dynamical decoupling sequence \cite{SouzaPRL2011} or a sequence from the universally robust family \cite{Genov2017PRL,ezzell2023dynamical}, which could, in principle, achieve even longer memory times. 
	%
	\begin{figure}[t!]
		\includegraphics[width=\columnwidth]{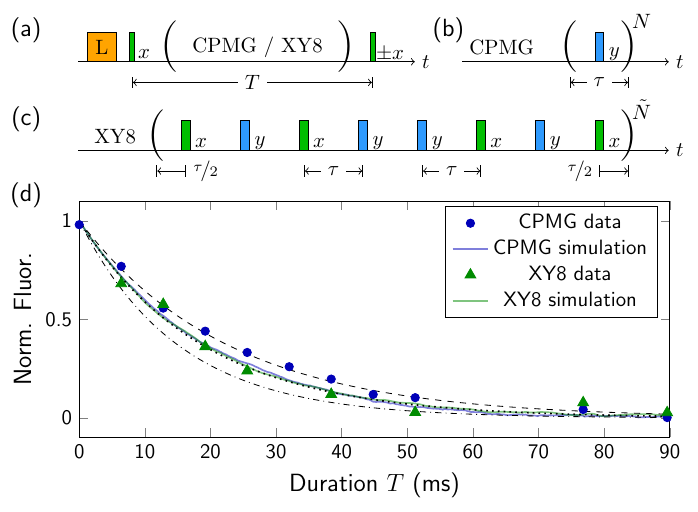}
		\caption{
			(a) Memory measurement sequence using the CPMG (b) and $XY8$ (c) protocol. Subscripts represent the phase of a pulse. (d) Experimental results for CPMG (blue dots) and $XY8$ (green triangles) with fixed $ \tau = 100\,\mu\text{s}$ and sweeping order $N$, $\tilde{N}$ vs. the total duration time $ T $. Fluorescence is normalized to the expected value for $T=0$ for each sequence \cite{SM}.
			Exponential decay fitting yields $T_{2,\text{CPMG}}= 24.1\pm0.9\,\text{ms}$ and  $ T_{2,XY8} = 18\pm3\,\text{ms}$.
			Solid lines represent OU simulations for CPMG (blue) and XY8 (green), closely matching the measured data. 
			The results remain within the simulation curves for the boundaries of the correlation time of 12.4 s (18.6 s), displayed as dashed (dash-dotted) lines (see the text and \cite{SM}).
		}
		\label{Fig:CPMG_XY8_N}
	\end{figure}
	\\Figure \ref{Fig:CPMG_XY8_N} shows the measurement sequences [Figs.~4(a)-4(c)] and the corresponding decay curves [Fig.~4(d)], where we progressively increase the memory time by changing the order $N$ or $\tilde{N}=N/8$ (for $XY8$), while keeping $\tau$ constant. 
	A simple exponential fit to the data yields $T_{2,\text{CPMG}}=24.1\pm 0.9\,$ms and $T_\text{2,XY8}=18\pm3\,$ms \cite{SM}. 	
	Compared to the Hahn echo $T_2$ this is a 45-fold (CPMG) and 40-fold ($XY8$) extension, respectively.	
	
	\textit{Noise model and numerical simulation.---}
	%
	In order to characterize the performance of the DD sequences and possibly prolong the coherence time we perform numerical simulations of decoherence during DD.
	For this purpose, we consider a simplified model of a two-state quantum system, which is subject to magnetic noise and power fluctuations of the driving fields.
	The decoherence model is similar to the one used in other color centers in diamond, e.g., NV centers \cite{DeLange2010,Aharon2016NJP,Genov2019MDD,Genov2019MDD,Aharon2016NJP,Genov2020PRR}. 
	It assumes that resonant interactions (flip flops) between the GeV and the bath spins (apart from $^{13}$C) are negligible due to a large energy mismatch. 
	Thus, the effect of the bath is dephasing of the GeV spin and can be approximated by magnetic noise along the GeV’s quantization axis. 
	In order to analyze decoherence during DD we consider the Hamiltonian in the rotating frame at the carrier frequency $\omega$ of the pulses after applying the rotating-wave approximation ($\Omega\ll\omega$) \cite{SM}
	\begin{align}\label{H_sensing_1s}
		H_{1}(t)=\frac{\delta(t)}{2}\sigma_{z}+\frac{\widetilde{\Omega}(t)}{2}\{&\cos{[\phi(t)]}\sigma_{x}+\sin{[\phi(t)]}\sigma_{y}\},
	\end{align}
	where $\widetilde{\Omega}(t)=\Omega[1+\epsilon(t)]f(t)$ is the magnitude of the Rabi frequency with $\Omega=(2\pi)\,6.486$\,MHz its target peak value, $f(t)$ describes its expected time dependence (e.g., it can be $0$ or $1$), $\epsilon(t)$ characterizes the amplitude noise, and $\phi(t)$ is its relative phase (e.g., $0^{\circ}$ or $90^{\circ}$). 
	The detuning $\delta(t)$ is the difference in the Larmor frequency of the GeV electron spin from the angular frequency of the driving field $\omega$, e.g., due to the hyperfine splitting and magnetic noise. 
	Similarly to other experiments in color centers in diamond \cite{DeLange2010,Aharon2016NJP,Genov2019MDD,Aharon2016NJP,Genov2020PRR}, we model $\delta(t)$ with an Ornstein-Uhlenbeck (OU) process \cite{UhlenbeckRMP1945,Gillespie1996AJP} with a zero expectation value $\langle \delta(t)\rangle = 0$ and correlation function $\langle \delta(t)\delta(t^{\prime})\rangle =\sigma_{\delta}^2\exp{(-\gamma|t-t^{\prime}|)}$, where $\sigma_{\delta}^2=\langle \delta(t)^2\rangle$ is the variance of the detuning due to noise, $\sigma_{\delta}\approx \sqrt{2}/T_{2}^{\ast}\approx 2\pi~146$ kHz \cite{Gillespie1996AJP,DeLange2010,Pascual-WinterPRB2012,SM}, with $T_2^{\ast}\approx 1.43~\mu$s the decay time of the signal from the Ramsey measurement in Fig. \ref{Fig:Spin_properties}(c). 
	We fit the decay shape of the signal from the CPMG and $XY8$ experiments and obtain an estimate of the correlation time $\tau_c=1/\gamma$ in the range of $12.4$ and $18.7$ s with an expected value of $\tau_c\approx 15.5 $ s \cite{SM}. The variation in the estimated values is likely due to fit uncertainty and slight changes in magnetic noise during operation, e.g., due to drift in temperature or alignment.
	We plot the theoretical coherence decay curves for DD with ideal, instantaneous $\pi$ pulses for an OU process \cite{SM,Pascual-WinterPRB2012} for the expected $\tau_c= 15.5\,$s (dotted line) and the upper (lower) values of the estimated range $\tau_c=18.7\,$s ($\tau_c=12.4\,$s) as dashed (dashed-dotted) lines in Fig. \ref{Fig:CPMG_XY8_N}(d).
	\\
	The amplitude error $\epsilon(t)$ is also modeled by an OU process with standard deviation  $\sigma_{\epsilon}=0.005$ and correlation time $\tau_{\Omega}=500\,\mu$s, similarly to previous work \cite{Aharon2016NJP,SM}.
	We calculate the  $\delta(t)$ and $\epsilon(t)$ for $2500$ different noise realizations, simulate the evolution of the system for each and obtain the average the density matrix from all noise realizations.
	The simulated signal decay \cite{SM} is shown as solid blue (CPMG) and solid green ($XY8$) lines in Fig. \ref{Fig:CPMG_XY8_N}(d), resulting in simulation estimates for the coherence times of $T_{2,\text{CPMG}}=19.8\,\text{ms}$ and $T_{2,XY8}=19.2\,\text{ms}$ \cite{SM}.
	The experimental data fit well to the simulation results, especially for $XY8$, while the CPMG data also lie within the expected range of the theoretical decay curves for the noise model. 
	The good fit of the experimental data, simulation results and theoretical decay curves indicate excellent control of the system and compensation of experimental imperfections. 
	CPMG slightly outperforms $XY8$, most likely due to the effect of spin locking and possibly a nonzero interaction of the GeV electron spin with the strongly coupled $^{13}$C for $XY8$ at $\tau=100\,\mu$s \cite{SM}, which is not considered in the simulations.
	The good agreement between experiment and simulation confirms the OU process as a valid way to model the environmental noise and field errors, identifying them as the main limit for the coherence time.
	The noise model can also be applicable to other color centers in diamond, e.g., SiV centers \cite{sukachev2017silicon}, to enhance the understanding of decoherence for group-IV defects.
	This, in principle, allows for the design of optimized control sequences to prolong the coherence time further, e.g., by carefully choosing the interpulse delay or using higher-order DD \cite{ajoy2011optimal,Genov2017PRL,ezzell2023dynamical}.
	
	\textit{Conclusion.---}
	%
	We demonstrated for the first time efficient initialization, readout and coherent control of the electron spin of the GeV at millikelvin temperatures.
	We applied dynamical decoupling sequences and increased the coherence time by several orders of magnitude to more than $ 20\,\text{ms}$, which is the longest coherence time for group-IV defects up to date, to the best of our knowledge. 
	The performed decoherence simulations fit the experimental data reasonably well, validating the noise model and allowing for the design of optimized control schemes for GeV and other group-IV defects.
	Using isotopically enriched $^{12}$C diamonds could allow for even longer memory times.
	Another strategy to enhance memory time involves 
	storing the quantum state in long-lived nuclear spins, either those inherent to GeV itself \cite{adambukulam2023hyperfine} or the neighboring $^{13}$C spins through dynamical decoupling applied to the GeV electron spin \cite{Nguyen2019NanophotonicInterface,Nguyen2019PhysRevB,MaityMechanicalControlNuclearSpin}.
	The results demonstrate the applicability of the GeV as a quantum memory, overcoming one main obstacle for quantum technology applications of group-IV defects, e.g., for quantum communication.

	\begin{acknowledgments}
		We thank Yuri N. Palyanov, Igor N. Kupriyanov, and Yuri M. Borzdov for providing the sample used in this work.
		This work was supported by DFG via Projects No. 386028944, No. 445243414, and No. 387073854, ERC Synergy grant HyperQ (Grant No. 856432), Baden-W\"urttemberg Stiftung and Volkswagen Stiftung, BMBF via project QuMicro, SPINNING, CoGeQ, QR.X, and Quantum HiFi. 
	\end{acknowledgments}
	
	\bibliographystyle{ieeetr}
	\bibliography{references}
	

	\pagebreak

	\renewcommand{\thefigure}{S.\arabic{figure}}
	
	
	\setcounter{figure}{0}


	%
	%
	%
	%
	
\onecolumngrid
\section*{Supplemental Material to ``Germanium Vacancy in Diamond Quantum Memory Exceeding 20 ms''}
\section{Numerical simulation}

\subsection{The System}

In order to characterize the performance of our dynamical decoupling sequences we consider a simplified model of a two-state quantum system within the lower ground states manifold of the GeV (see Fig. 1 in the main text). Specifically, we treat the splitting due to strong coupling to nearby $^{13}$C as detuning, which is subject to magnetic noise. In addition, we model the  power fluctuations of the driving fields and simulate the system evolution numerically. 
The model is similar to the one used for modeling decoherence of other color centers, e.g., the NV center in diamond \cite{DeLange2010,Aharon2016NJP,Genov2019MDD}. It assumes that the GeV interacts with a a spin bath, where resonant interactions (flip-flops) between the GeV and the bath spins are minimum due to a large energy mismatch. Thus, the effect of the bath is dephasing of the GeV spin and can be approximated by magnetic noise along the GeV’s quantization axis, which leads to a time-dependent detuning $\delta(t)$ of the Larmor frequency from its expected value. 
Specifically, we consider the Hamiltonian
\begin{align}
	H(t)=&\frac{\omega_0+\delta(t)}{2}\sigma_{z}+\Omega_1 f(t)(1+\epsilon(t))\cos{(\omega_0 t+\phi(t))}\sigma_{x}),
\end{align}
where $\omega_0$ is the expected Larmor frequency of the GeV electron spin, $\Omega_1$ is the peak Rabi frequency of the driving field, $f(t)$ characterizes target variation of the Rabi frequency in time, e.g., it can be a step function taking values $0$ and $1$, while $\phi(t)$ is the phase of the applied field. 
The parameters $\delta(t)$ and $\epsilon(t)$ characterize the time-varying errors in the Larmor frequency of the GeV electron spin and the target Rabi frequency.

We then move to the interaction basis with respect to $H_0^{(1)}=\omega_0\sigma_{z}/2$ and obtain after applying the rotating-wave approximation ($\Omega_1\ll\omega_0$)
\begin{align}\label{H_sensing_1s}
	H_{1}(t)=\frac{\delta(t)}{2}\sigma_{z}+\frac{\Omega_1}{2}f(t)(1+\epsilon(t))(&\cos{(\phi(t))}\sigma_{x}+\sin{(\phi(t))}\sigma_{y}).
\end{align}
We consider this Hamiltonian in the simulation as the rotating-wave approximation is usually satisfied very well for our experimental parameters.

\subsection{Noise Model}

We use a noise model of the environment that has the characteristics for typical experiments in color centers in diamond \cite{Genov2019MDD,Aharon2016NJP,Genov2020PRR}. 
Specifically, the noise $\delta(t)$ is modelled as an Ornstein-Uhlenbeck (OU) process \cite{UhlenbeckRMP1945,Gillespie1996AJP} with a zero expectation value $\langle \delta(t)\rangle = 0$, correlation function $\langle \delta(t)\delta(t^{\prime})\rangle =\sigma_{\delta}^2\exp{(-\gamma|t-t^{\prime}|)}$, where $\sigma_{\delta}^2=\langle \delta(t)^2\rangle$ is the variance of the detuning due to noise, which is characterized by the standard deviation $\sigma_{\delta}=\sqrt{D\tau_c/2}$ with $D$ a diffusion constant and $\tau_c=1/\gamma$ the correlation time of the noise for the OU process \cite{Gillespie1996AJP}. The OU process is implemented with an exact algorithm \cite{Gillespie1996AJP}
\begin{equation}\label{Eq:OU_noise}
	\delta(t+\Delta t)=\delta(t)e^{-\frac{\Delta t}{\tau_c}}+\widetilde{n}_{\delta}\sqrt{\sigma_{\delta}^2\left(1-e^{-\frac{2\Delta t}{\tau_c}}\right)},
\end{equation}
where $\widetilde{n}_{\delta}$ is a unit Gaussian random number. 

We calibrate the effect of environmental noise from the decay time $T_2^{\ast}\approx 1.425~\mu$s of the signal from a Ramsey measurement (see Fig. 2(c) in the main text), the decay time $T_2=440~\mu$s of a spin echo measurement, where a $\pi$ pulse is applied in the middle of the interaction (see Fig. 2(d) in the main text), and the decay times of the CPMG sequences in Fig. 4 in the main text. 
We note that our spin echo measurement consists of a single block $\tau/2-\pi-\tau/2$, where $\tau/2$ is the free evolution before and after the refocusing $\pi$ pulse to correspond to the other dynamical decoupling sequences we use. In the literature spin echo measurements are typically performed with the sequence $\tau-\pi-\tau$, so the corresponding spin echo decay time will be $\widetilde{T}_2=T_2/2=220~\mu$s.

\begin{figure}[t!]
	\includegraphics[width=0.89\columnwidth]{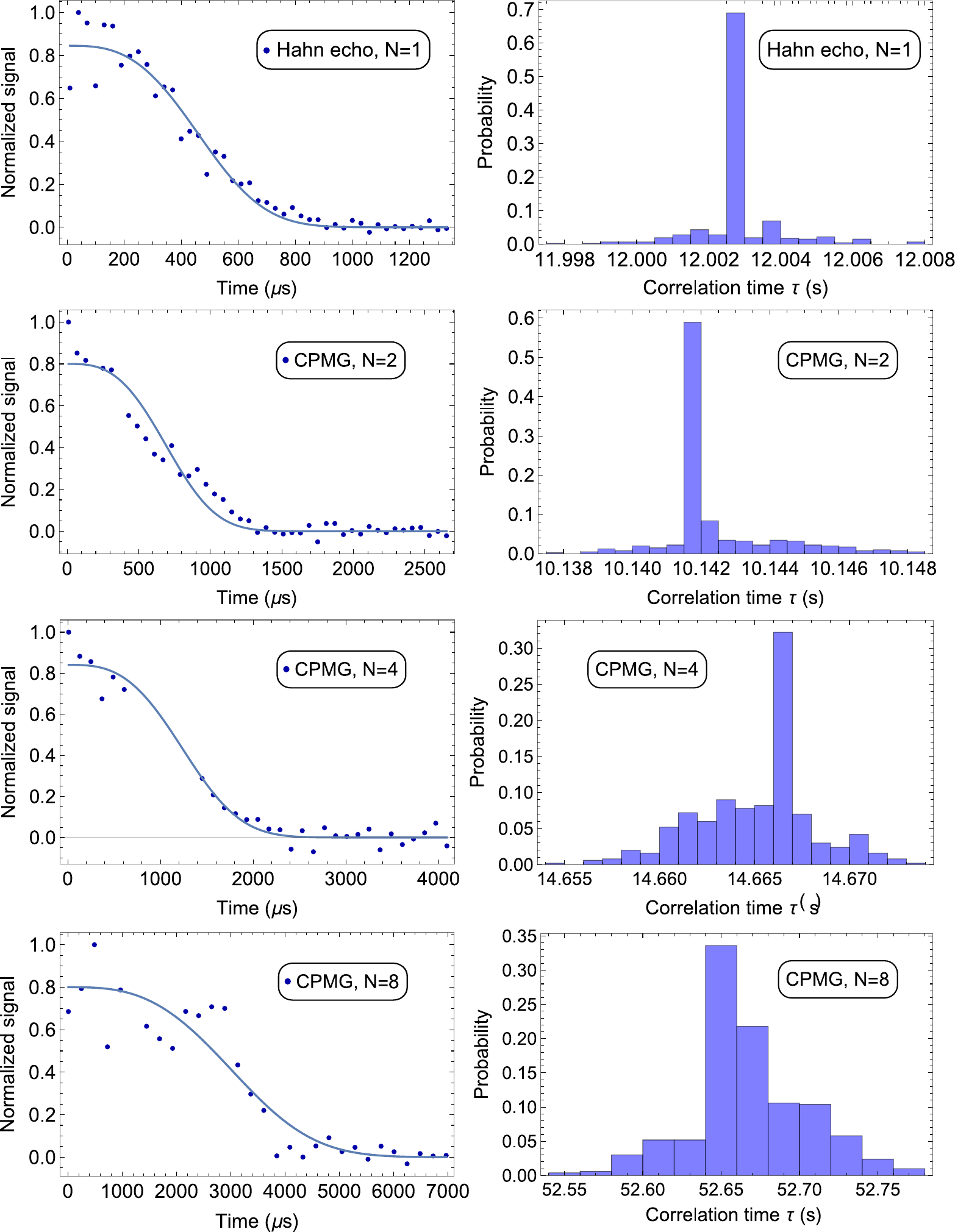}  	    
	\caption{
		Estimates of the correlation time $\tau_c$ for different CPMG sequences from Fig. 3 in the main text. The left figures show the experimental data and the best fit to the model in Eq. \eqref{Eq:OU_formula_full}. The right figures show the corresponding histograms of the estimated value of the correlation time $\tau_c$, obtained when performing the fitting 500 times, starting from random initial guesses for the experimental parameters (see text).
		The estimated correlation times from the best of all fits with random initial guesses, i.e., having maximum $R^2$, are as follows: 
		(a) $\tau_c(N=1)\approx 12\pm 0.96\,$s, 
		(b) $\tau_c(N=2)\approx 10.14\pm 0.97\,$s, 
		(c) $\tau_c(N=4)\approx 14.67\pm 1.42\,$s, 
		(d) $\tau_c(N=8)\approx 52.65\pm 8.53\,$s. 
	}
	\label{Fig:CPMG_tau_estimation}
\end{figure}

\begin{figure}[t!]
	\includegraphics[width=0.9\columnwidth]{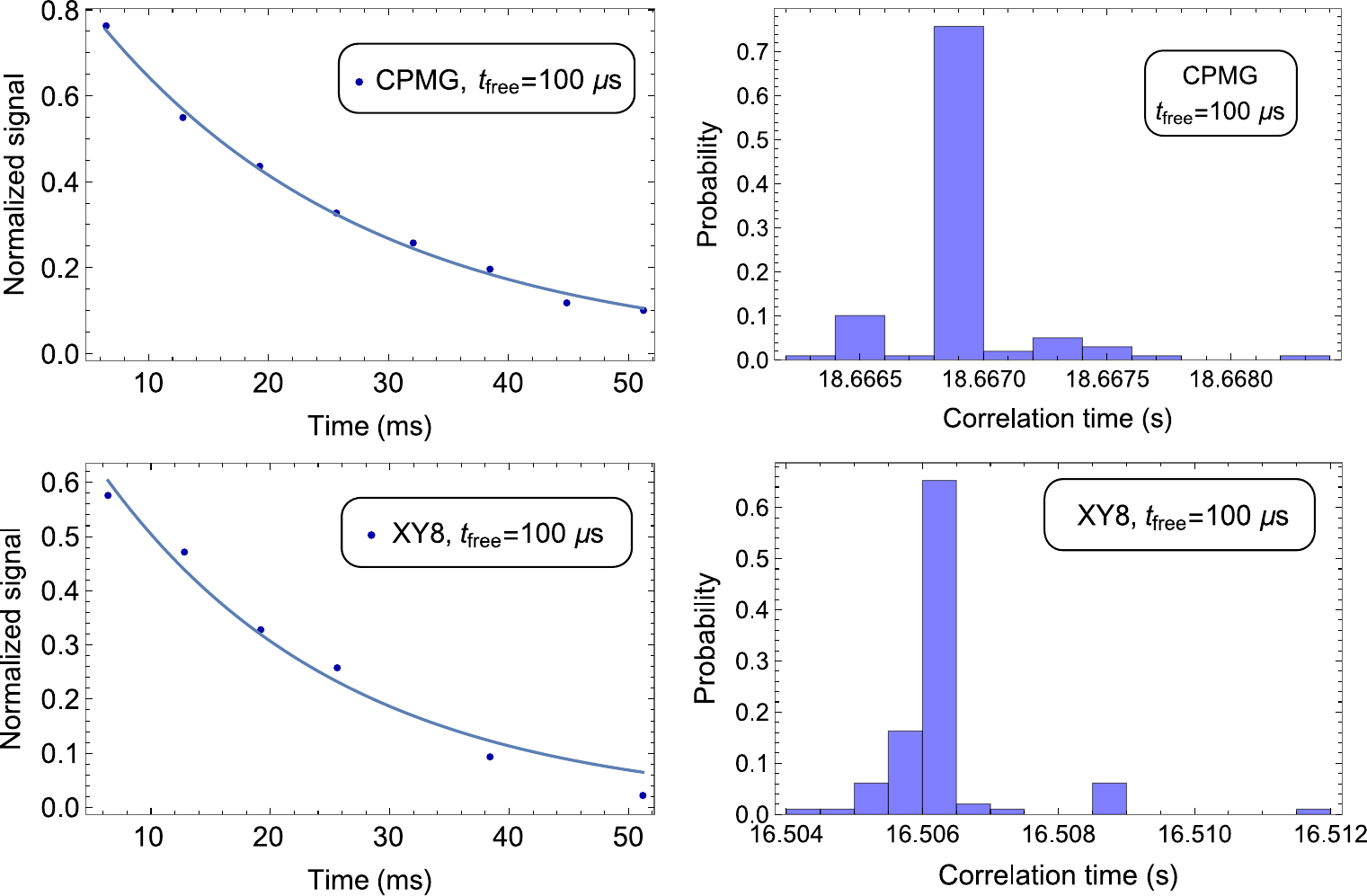}  	    
	\caption{
		Estimates of the correlation time $\tau_c$ for the CPMG and XY8 order scan sequences for $\tau=100\,\mu$s from Fig. 4 in the main text. The left figures show the experimental data and the best fit to the model in Eq. \eqref{Eq:OU_formula_full}. The right figures show the corresponding histograms of the estimated value of the correlation time $\tau_c$, obtained when performing the fitting 50 times, starting from random initial guesses for the experimental parameters (see text). The estimated correlation times from the best of all fits with random initial guesses, i.e., having maximum $R^2$, are as follows: (a) $\tau_c\,(\text{CPMG}, t_{\text{free}}=100\,\mu s)\approx 18.67\pm 0.53\,$s, (b) $\tau_c\,(\text{XY8},t_{\text{free}}=100\,\mu s)\approx 16.51\pm 1.8\,$s. 
	}
	\label{Fig:DD_N_fits}
\end{figure}

In the limit of long correlation time $\tau_c\gg T_{2}^{\ast}$, as in our case, the value of $T_{2}^{\ast}\approx 1.425\,\mu$s is determined mainly by $\sigma_{\delta}$, allowing us to obtain  $\sigma_{\delta}\approx \sqrt{2}/T_{2}^{\ast}\approx 2\pi~146$ kHz \cite{Gillespie1996AJP,DeLange2010,Pascual-WinterPRB2012}. 
We then obtain the correlation time $\tau_c$ by fitting the exact analytical formula for the expected coherence decay rate $\gamma(N,\tau)$ due to magnetic noise, modelled with an OU process, during a sequence of $N$ ideal, instantaneous $\pi$ pulses for total evolution time $t=N \tau$ \cite{Pascual-WinterPRB2012}:
\begin{equation}\label{Eq:OU_formula_full}
	\gamma(N,\tau)=\sigma _{\delta }^2\tau_c ^2  \left[-\left((-1)^{\text{N}+1} e^{-\frac{t}{\tau_c}}+1\right)
	\left(1-\text{sech}\left(\frac{\tau}{2 \tau_c }\right)\right)^2+t
	\left(\frac{1}{\tau_c }-\frac{2 \tanh \left(\frac{\tau}{2 \tau_c
		}\right)}{\tau}\right)\right],
\end{equation}
where $t=N \tau$. In the simplest case of a Hach echo, where $N=1$ and in the limit of $t=\tau\gg \tau_c$, the Hahn echo decay rate simplifies to \cite{Pascual-WinterPRB2012}
\begin{equation}\label{Eq:OU_formula_SE}
	\gamma_{\text{se}}(\tau)=\sigma _{\delta }^2\tau_c t.
\end{equation}
Then, the relation between Hahn echo $T_{2}$ and the OU noise parameters is given by $T_{2}\approx 2(3/D)^{1/3}$. The latter formula allows in principle to obtain $D$ and the correlation time $\tau_c\approx 4/(T_{2}^{2\ast} D)$. While this is estimation procedure is usually sufficient, we use the full formula in Eq. \eqref{Eq:OU_formula_full} for fitting to obtain a more precise value of the correlation time and reduce inaccuracy due to approximations. In addition, we fit the OU model not only the Hahn echo data but also to the experimental results for the dynamical decoupling sequences in Figs. 3 and 4 in the main text. 

Figure \ref{Fig:CPMG_tau_estimation} shows the fits of the model to the data in Fig. 3 in the main text, where we vary the free evolution time $\tau$ for several different CPMG sequences. We used the NonLinearModelFit procedure in Mathematica for fitting the model in Eq. \eqref{Eq:OU_formula_full} to the data. In addition, we fixed the estimate of $\sigma_{\delta }\approx \sqrt{2}/T_{2}^{\ast}\approx 2\pi~146$ kHz. The best fits and the corresponding experimental data are shown in the left column of Fig. \ref{Fig:CPMG_tau_estimation} for each of the experiments. We estimate the correlation time by running the estimation procedure 500 times for each of the experimental datasets, starting from random initial guesses for the model parameters. The histograms of the estimated correlation times are shown in the right columns. The same procedure is performed in Fig. \ref{Fig:DD_N_fits} for the CPMG and XY8 order scans in Fig. 4 in the main text. 

We estimate $\tau_c$ in the range of 12-19 s for all sequences, except for CPMG-2 (most likely due to effects of nearby $^{13}$C spins for the particular sampled free evolution times) and CPMG-8 (most likely due to a change in experimental conditions), which produce outliers. 
We note that we removed some of the data points for CPMG-4 around $\tau_c=1$ ms as there was a significant drop to the signal, again attributed to $^{13}$C interaction. We estimate $\tau_c\approx 15.5$ s, calculating as its average estimated value of $\tau_c$ from all experiments, except the outliers for CPMG-2 and CPMG-8. We use this estimate of $\tau_c$ in our simulations in the main text.

\begin{figure}[t!]
	\includegraphics[width=0.7\columnwidth]{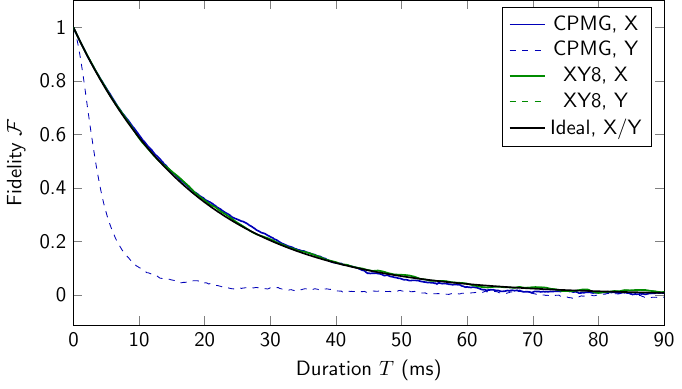}
	\caption{Simulation of the fidelity of the CPMG and XY8 sequences for different initial states (see Fig. 4 in the main text). The initial state X is obtained after a first $\pi/2(y)$ pulse (with $\phi(t)=90^{\circ}$, see Eq. \eqref{H_sensing_1s} in this Supplemental material) and Y -- after a $\pi/2(x)$ pulse (with $\phi(t)=0^{\circ}$. We vary the total measurement time $T=N\tau$ by chaning the order $N$ and $\widetilde{N}=N/8$ of CPMG and XY8 and keeping $\tau_c=100\,\mu$s. It is evident that fidelity of CPMG with an initial Y state is much worse than the one of the X state and the fidelities of all the other sequences. In contrast, the performance of XY8 does not depend on the initial state and is close to the one with ideal instantaneous $\pi$ pulses. }
	\label{Fig:CPMG_XY8_N_X_Y}
\end{figure}

Apart from magnetic field variation, we also model the fluctuations in the driving field Rabi frequency. Its errors are mainly determined by the experimental characteristics our arbitrary waveform generator and amplifier, so we assume that their noise characteristics are similar to the ones from previous experiments \cite{AharonPRL2018,Genov2019MDD,Genov2020PRR}. Specifically, we model the relative error of the Rabi frequency with an OU process with the update function \cite{Genov2019MDD}
\begin{equation}\label{Eq:OU_noise_epsilon}
	\epsilon(t+\Delta t)=\epsilon(t)e^{-\frac{\Delta t}{\tau_{\Omega}}}+\widetilde{n_{\epsilon}}\sqrt{\sigma_{\epsilon}^2\left(1-e^{-\frac{2\Delta t}{\tau_{\Omega}}}\right)},
\end{equation}
where $\sigma_{\epsilon}=\sqrt{(1/2)D_{\Omega}\tau_{\Omega}}=0.005$, the correlation time $\tau_{\Omega}=500 \mu$s with the corresponding diffusion constant $D_{\Omega}=2\sigma_{\epsilon}^2/\tau_{\Omega}$ \cite{Aharon2016NJP}.
As already evident, the initial values of $\delta(t=0)$ and $\epsilon(t=0)$ change from run to run and are taken from a Gaussian distribution with standard deviations $\sigma_{\delta}$ and $\sigma_{\epsilon}$, respectively.

\subsection{Fidelity Calculation}

We calculate the update functions for $\delta(t)$ and $\epsilon(t)$ for the particular run and calculate numerically the propagator
\begin{equation}\label{U_numeric}
	U_{1}(t,0)=\mathcal{T} \exp{\left(-i\int_{0}^{t}\widetilde{H}_{1}(t^{\prime})d t^{\prime}\right)},
\end{equation}
for the particular noise realisation of $\delta(t)$ and $\epsilon(t)$ and the chosen time dependence of $f(t)$ with $\mathcal{T}$ a time-ordering operator. We use a time-discretization with a time step of $\Delta t = 0.05$ ns when we apply the refocusing $\pi$ pulses and $\Delta t = 25$ ns during free evolution between the pulses. The former is comparable to the resolution of standard arbitrary wave-form generators. The reason for the difference in time discretization during and between the pulses is to optimize the simulation speed as the $\pi$ pulses are much shorter than the free evolution times between them. We note that the OU noise characteristics are not affected by this choice of $\Delta t$, as Eqs. \eqref{Eq:OU_noise} and \eqref{Eq:OU_noise_epsilon} are exact \cite{Gillespie1996AJP}.

\begin{figure}
	\includegraphics[width=\textwidth]{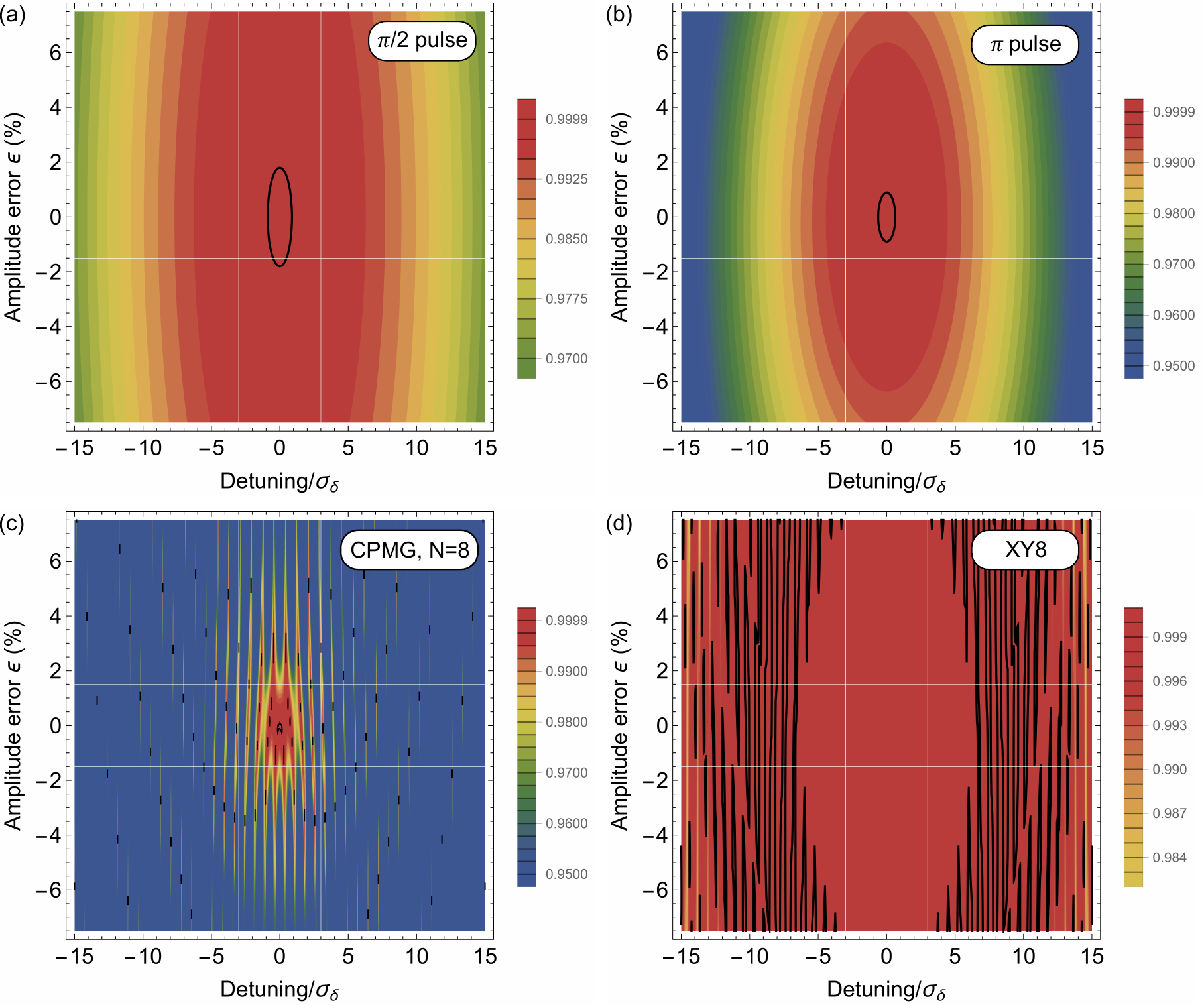}  	    
	\caption{
		Simulation of gate fidelity vs. amplitude and detuning errors of (a) a $\pi/2$, (b) a $\pi$ pulse, (c) the CPMG sequence, repeated to include a total of $N=8$ pulses, (d) the XY8 sequence. The time separation between the centers of the $\pi$ pulses of the CPMG and XY8 sequence is 100 $\mu$s, similarly to the experiment. The black lines indicate the typical quantum information threshold of $F_{\text{gate}}(t)=0.9999$. The white lines indicate the range of $\pm 3$ times the standard deviations of the amplitude ($\sigma_{\epsilon}=0.005=0.5\%$) and frequency detuning ($\sigma_{\delta }\approx \sqrt{2}/T_{2}^{\ast}\approx 2\pi~146$ kHz), used in the simulations. It is evident that XY8 compensates the expected errors very well and can be used for quantum memory applications.
	}
	\label{Fig:Gate_fidelity}
\end{figure}

We then make use of the calculated $U_{1}(t,0)$ and obtain the time evolution of the density matrix
\begin{equation}
	\rho(t)=U_{1}(t,0)\rho(0)U_{1}^\dagger(t,0),
\end{equation}
where $\rho(0)$ is the initial density matrix for the sensing protocol, e.g., $\rho(0)=\rho_{x}\equiv(I+\sigma_{x})/2$ when the system is initially along the $x$ axis of the Bloch sphere. We note that $I$ is the identity matrix and $\sigma_k$, $k=x,y,z$ are the respective Pauli matrices. Then, we calculate the expected density matrix $\overline{\rho}(t)=\frac{1}{n}\sum_{k=1}^{n}\rho_{k}(t)$ by averaging the density matrices $\rho_{k}(t),k=1\dots 2500$ for $n=2500$ noise realizations. We define the perfect final density matrix $\rho_{\text{no noise}}(t)=U_{1,\text{no noise}}(t,0)\rho(0)U_{1,\text{no noise}}^\dagger(t,0)$ when there is no noise, i.e., $\delta(t)=\epsilon(t)=0$ and calculate the fidelity for the particular initial density matrix $\rho(0)$ 
\begin{equation}
	F_{\rho(0)}(t)=\text{Tr} \left(\rho_{\text{no noise}}(t)\overline{\rho}(t)\right).
\end{equation}
We note that the performance of some DD sequences varies for different initial states. For example, Figure \ref{Fig:CPMG_XY8_N_X_Y} shows that CPMG with an initial X state, i.e., $\rho(0)=\rho_{x}$, performs similarly to an ideal sequence with perfect, instantaneous $\pi$ pulses even in the presence of pulse errors and magnetic noise. Then, the $T_{2,\text{,CPMG,X}}\approx 19.2$ ms for a pulse separation $\tau=100\,\mu$s. The reasons is that for this particular initial state the effect of a pulse error during an odd pulse is approximately compensated by the error of its subsequent even pulse \cite{carr1954effects,MeiboomGill1960}. It has been shown that then CPMG is effectively equivalent to spin locking and its coherence time converges to $T_{1,\rho}$ \cite{Santyr1988JMR}. However, the fidelity of its Y state, i.e., $\rho(0)=\rho_{y}$, is much worse with $T_{2,\text{CPMG,Y}}\approx 4.7$ ms as then the errors accumulate. 
In contrast, the performance of the XY8 sequences is identical for the X and Y initial states with $T_{2,\text{XY8,X}}\approx T_{2,\text{XY8,Y}}\approx 19.2$ ms and is close to the ideal one. This is due to the particular choice of the relative phases that allow for pulse errors compensation for arbitrary initial states. 
As the coherence time of CPMG in our experiment should theoretically approach $T_{1,\rho}$, it is expectedly slighly higher than the one of XY8.

Finally, we characterize the gate fidelities of our $\pi/2$, $\pi$ pulses, as well as the CPMG and XY8 sequences. In order to do this we define the gate fidelity similarly to \cite{Souza2012,Genov2017PRL} as 
\begin{equation}
	F_{\text{gate}}(t)=\frac{1}{2}\left|\text{Tr} \left(U_{1,\text{no noise}}^\dagger(t,0)U_{1}(t,0)\right)\right|,
\end{equation}
where $U_{1,\text{no noise}}(t,0)$ is the target propagator of the gate without noise and $U_{1}(t,0)$ is the actual one. As the correlation time of the noise is much longer than the duration of the $\pi/2$ and $\pi$ pulses, as well as the CPMG and XY8 sequences, we assume for simplicity that the frequency and amplitude noise terms $\delta(t)$ and $\epsilon(t)$ are constant during the interaction. 
Figure \ref{Fig:Gate_fidelity} shows a simulation of the gate fidelity of each of the techniques vs. amplitude and detuning errors. The simulation shows that the $\pi/2$ and $\pi$ pulses have quite high fidelity (Fig. \ref{Fig:Gate_fidelity}a,b). However, the gate errors accumulate for CPMG, e.g., when it is repeated and consists of eight $\pi$ pulses (Fig. \ref{Fig:Gate_fidelity}c). On the contrary, the errors are compensated very well for XY8 for the same number of pulses (Fig. \ref{Fig:Gate_fidelity}d), demonstrating its robustness.  
In addition, the experimental coherence time of XY8 is very close to the one of CPMG for its preferred state  and to the theoretical decay from the OU process with ideal, instantaneous pulses. This confirms that pulse errors are compensated efficiently in the experiment with the XY8 sequence. We note that even more advanced sequences can be applied like the Knill dynamical decoupling (KDD) sequence \cite{SouzaPRL2011} or a sequence from the universally robust (UR) family \cite{Genov2017PRL,ezzell2023dynamical}, which provide a higher order of error compensation in comparison to XY8 and could in principle achieve even longer memory times.

\newpage
\section{Dilution Refrigerator Setup}

\begin{figure}[t!]
	\includegraphics[height=0.39\textwidth]{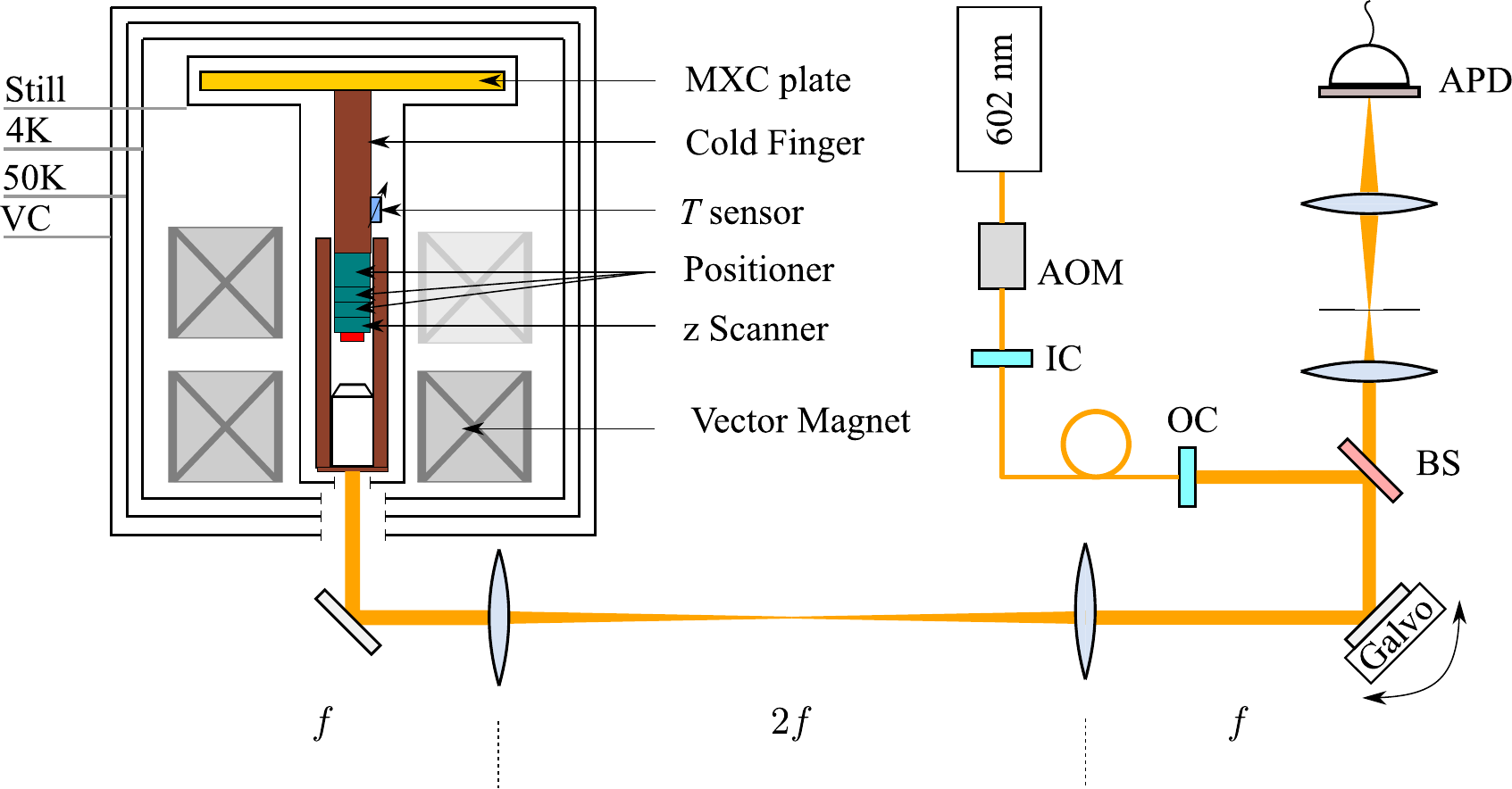}
	\caption{
		Optical experimental setup. 
		Sketch of dilution refrigerator (not to scale) with optical access from the bottom through windows.
		Integration into a home-built 4$f$ scanning confocal microscope for optical excitation and detection.
		The insert includes a cold finger attached to the MXC plate and a clamped objective holder.
		A 1T/1T/3T superconducting vector magnet operated in persistent switch mode is used.
		Temperature was monitored using a Lakeshore sensor at the cold finger.
		Sample positioning employed Attocube positioners and a z-scanner.
		Optical excitation and collection utilized a room temperature objective.
		Resonant addressing of the color center is achieved using a tunable CW dye laser. 
		Pulsing is implemented via acousto optical modulator (AOM).
		Subsequential spatial mode cleaning is done over coupling (IC: incoupler, OC: outcoupler) to a fiber.
		Lateral scanning was performed externally by stirring a mirror in a 4$f$ system. 
		Excitation and detection were seperated by beam sampler (BS).
		Fluorescence was collected in the phonon side band and detected with a single photon detector (APD). 
		Further details can be found in the text.
	}
	\label{Fig:Exp_setup}
\end{figure}	
The experiments are performed in a cryogen free dilution refrigerator from Bluefors (BF-LD400) with a base temperature of 13\,mK (loaded) at the mixing chamber (MXC).
The cryostat allows for free optical access from the bottom through windows installed in the different shielding layers.
For optical excitation and detection of the GeV the cryostat is implemented in a home-built 4$f$ scanning confocal microscope (Fig. \ref{Fig:Exp_setup}).
We use a single frequency tunable CW dye laser Matisse DS from Sirah Lasertechnik operated at a central wavelength of 602\,nm (Rhodamin B in ethylene glycol)  for resonant addressing of the color center.
Laser pulses with a rise time of less than 20\,ns are realized using an acousto optic modulator (AOM) from Crystal Technology (Model 3200-146) with a self-built AOM-driver.
By coupling the light into a bare photonic crystal fiber (core diameter 5\,$\mu$m) provided by NKT Photonics, the mode is spatially cleaned from artifacts caused by the AOM.
An Olympus Plan Achromat Objective type RMS10X is used for incoupling (IC) and type RSM4X for outcoupling (OC).
We choose the OC such that the collimated beam size is overfilling the back aperature of the objective used within the confocal microscope.
This allows us to cope for imperfections in the alignment arising during the cool down of the cryostat without loosing our signal.
This beam is then guided to the dilution refrigerator.

The confocal microscope insert consists of mainly two parts, the cold finger on which the sample is mounted and the objective holder.
The cold finger is realized with a $\sim$30\,cm long solid copper rod attached to the MXC plate.
At the end, a U-shaped copper support structure with RMS thread is clamped to the cold finger to serve as a mounting platform for the objective lens, facing the sample.
This ensures the proper thermal grounding of the objective to the milikelvin stage.
This configuration reaches into the bore of a 1T/1T/3T (x/y/z) superconducting vector magnet from American Magnetics which is attached to the 4\,K stage of the cryostat.
The magnet is operated in persistent switch mode to minimize heat load and magnetic field fluctuations.

For the initial positioning of the sample a xyz-slip stick positioner stack (Attocube ANPx101/z102/RES, titanium) with resistive feedback is installed on the cold finger.
Additionally, we use an Attocube copper beryllium z-scanner ANSz100std allowing for depth scans of the sample.
The sample is mounted via indium soldering on our self-made copper sample holder.
To ensure a good thermal grounding an attocube ATC100 plate between the sample holder and the cold finger is used.
The temperature is monitored with a temperature sensor (Lakeshore RX-102A) mounted on the cold finger, as can be seen in Fig. \ref{Fig:Exp_setup}.
Despite the implemented thermal grounding, the temperature of the sample could differ to the one shown on the sensor due to the different positions.

For optical excitation and collection we use a Newport objective model LI-60X with a NA of 0.85, which is mounted within the cryostat and thermalized at the milikelvin stage, as previously described. This objective is not rated for cryogenic application. However, from the author's experience there is no indication for significant difference to the performance of a objective rated for cryogenic operation.
To minimize the heat load the lateral scanning is performed by changing the angle of the incident beam on the back aperture of the objective.
This is realized outside of the cryostat using a 4$f$ system with a Piezo tip/tilt platform (PI S-335.2SH).
The fluorescence is collected in the phonon side band (PSB) using a beam sampler (Thorlabs BSF20-B) and a bandpass filter in front of the single photon detector (Excelitas SPCM-780-44).
This implementation of the confocal setup allows for real time adjustments and modifications of the setup with respect to the needs.
Moreover, as most of the optical components are placed outside the cryostat the heat load is reduced.

The pulsed measurement are performed with an arbitrary wavefunction generator (AWG) from Tektronix (model 70000B HP).
The generated microwave pulses are amplified with an ar amplfier (model 50S1G6) and then delivered via semirigid coaxial cables to the cryostat.
Inside the cryostat the MW is provided through semirigid coaxial cables type 2.19mm SCuNi-CuNi from the RT to the 4\,K plate and type 0.86mm SCuNi-CuNi fromt he 4\,K to the MXC plate.
As already written in the main text, the MW was delivered by a 20\,$\mu$m thick copper wire, soldered onto a self-designed PCB.
The PCB is soldered to semiridig coaxial cables, which are connected to the MXC plate.
The poor soldering connection points and not impedance matched PCB is one of the main sources of heat load during coherent control with MW.
This can be certainly improved by using superconducting lines towards and on the sample. 
%

\section{Supplementary information on GeV physical system}

\begin{figure}[t!]
	\includegraphics[width=0.9\textwidth]{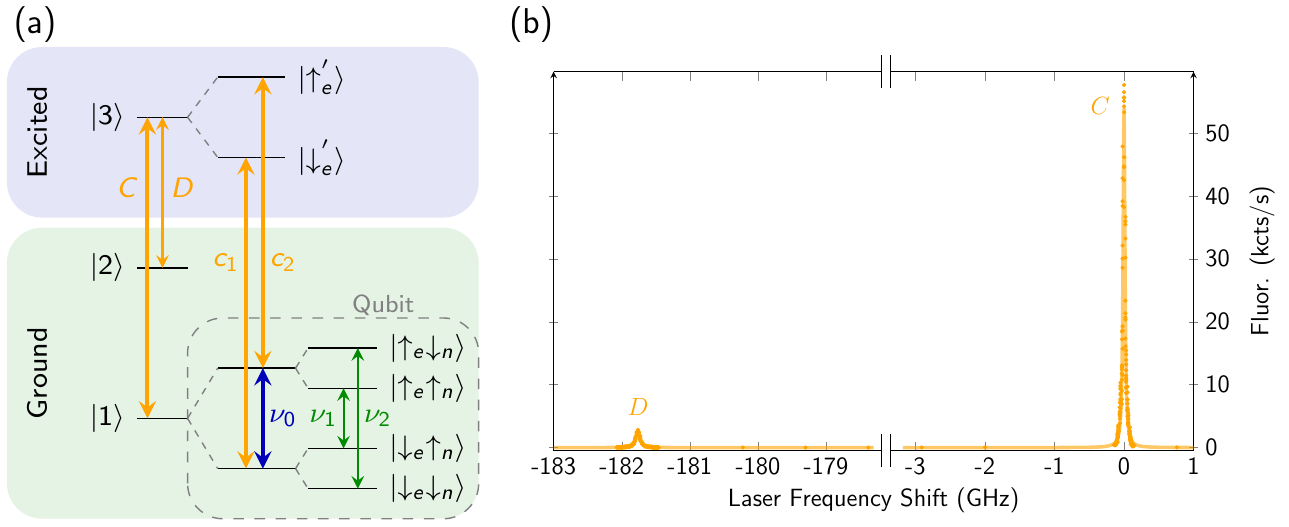}
	\caption{
		(a) Extended level scheme of Fig. 1(a) of the GeV center under investigation, which includes ground state $|2\rangle$.
		The frequency difference between levels $|1\rangle$ and $|2\rangle$ is not up to scale (see the main text for details).
		(b) PLE measurement of the transitions $C$ and $D$, taken at 4\,K show a splitting of 181.760\,GHz, indicating strain.
	}
	\label{Fig:PLE_CD}
\end{figure}

In the main text, we have discussed the main characteristics of the investigated GeV. In this section, we aim to provide additional details for the sake of completeness and to improve presentation. 
Figure \ref{Fig:PLE_CD} (a) illustrates the level scheme of the GeV, including the upper ground state labeled as level $|2\rangle$. 
For the sake of simplicity, we still omit the upper excited state since, at the operating temperatures, it does not contribute to the spin dynamics. The ground state splitting ($\Delta_g$) mentioned in the main text corresponds to the splitting between level $|1\rangle$ and level $|2\rangle$. 
The representation of the levels is not to scale, particularly concerning the qubit sublevels in relation to levels $|1\rangle$, $|2\rangle$ and $|3\rangle$.  
In order to investigate our ground state splitting we perform  photoluminescence excitation (PLE) measurements at 4\,K.
By substracting the found resonances $C$ (transition between $|1\rangle$ and $|3\rangle$) and $D$ (transistion between $|2\rangle$ and $|3\rangle$) we can determine $\Delta_g$ (transition between $|1\rangle$ and $|2\rangle$) of our system to be approximately 181 GHz. 
By comparing the observed ground state splitting to previously reported ones \cite{siyushev2017gev,Bhaskar_PRL2017}, the GeV used in this work features a strain induced additional ground state splitting of $\approx$ 20\,GHz. 
As described in the main text, this allows to drive the microwave transitions $\nu_{\{0,1,2\}}$ (as shown in Fig. \ref{Fig:PLE_CD}), as this is partly lifting the orthogonality of the orbital states.\\

\begin{figure}[t!]
	\includegraphics[height=0.35\textheight]{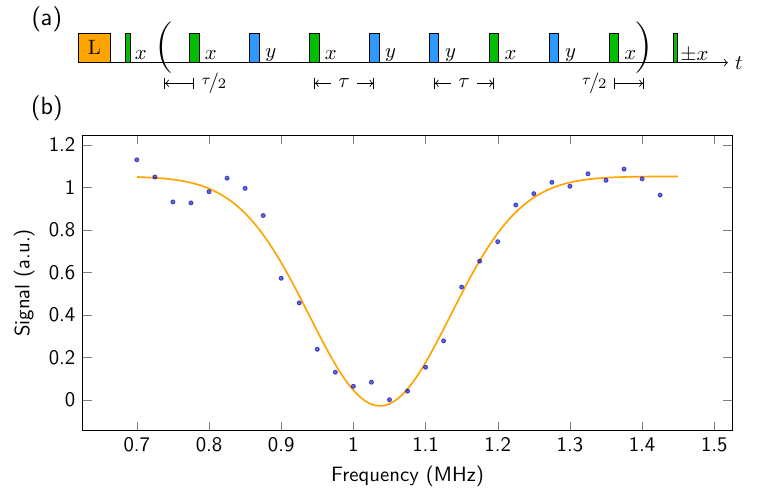}
	\caption{
		(a) Pulse scheme for alternating XY8-1 measurement.
		The variation in the frequency $f$ is performed by changing the interpulse duration $\tau$, using $f=\nicefrac{1}{2\tau}$
		(b) Alternating measured signal shows a dip at $1.036\pm0.003\,$MHz corresponding to the Lamor frequency of the surrounding $^{13}$C bath for the applied magnetic field.
	}
	\label{Fig:xy8_freq}
\end{figure}
To gain a comprehensive understanding of our system and the measurements we conducted, it is crucial to investigate not only the properties of the electron spin but also those of the surrounding spin bath. 
One significant contribution to the spin bath arises from the surrounding $^{13}$C isotopes with their spin $\nicefrac{1}{2}$ nature, which can have influence of the system.
To identify its associated frequencies, we performed an XY8-1 measurement (as shown in Fig. \ref{Fig:xy8_freq}(a)), where we systematically varied the interpulse duration $\tau$. 
As can be seen by the depicted sequence, the measurement was performed in an alternating manner. By changing the phase of the latter $\pi/2$-pulse between x ($0^\circ$) and -x ($180^\circ$) the read out is performed either in dark state or in the flipped state.
The subtraction of the signals from both readouts results then in the normalized, differential signal, which takes into account laser fluctuations for the measurement sequence.
Fig. \ref{Fig:xy8_freq} (b) illustrates the measured differential signal in relation to the mapped frequencies. 
By analyzing this data, we were able to determine the Larmor frequency of 1.036 MHz for the applied magnetic field.
This finding provides insights into the coupling mechanisms and interactions between the electron spin and the surrounding environment and helps to enhance the understanding of the measurement results. 
By identifying the specific frequency associated with the spin bath, we can better distinguish its effects from other factors influencing the electron spin dynamics in our system.

\section{Additional details about the memory time experiments}
%
The memory time of a quantum network node plays a crucial role in quantum information processing (QIP), as it determines the duration for which information can be stored and manipulated in quantum-mechanical states.
Therefore, it is of great significance to identify sequences that preserve the quantum state and thereby extend the memory time.
As already described in the main text, for memory type experiments with dynamical decoupling sequences a specific $\tau$ is chosen and the order $N$ is varied \cite{ajoy2011optimal,SouzaPRL2011,Pascual-WinterPRB2012,Bar-GillNatComm2013,Zhong2015,Genov2017PRL,ezzell2023dynamical}. 
For this study, we conducted measurements using both, CPMG and XY8 types of dynamical decoupling, mainly consisting out of equally spaced $\pi$ pulses, where we used their relative phases to compensate pulse errors and refocus the quantum state of the GeV center (c.f. main text). 
In our experiment these $\pi$ pulses are realized using MW control, which can introduce significant heat load to the experimental setup.
Due to the limited cooling power of our dilution refrigerator we adjusted carefully the pulse sequence, in terms of the choice of a favorable $\pi$ pulse spacing $\tau$ and in terms of reduction of the the duty cycle as an established technique for heating mitigation.
For our experiment, we have thus chosen a pulse spacing of $\tau=100\,\mu$s for the memory measurements, as the Hahn echo shows a negligible decay for this time and the accumulated heating of the pulses was still tolerable. 
In particular for the memory measurements, we reduce the duty cycle of the overall pulse sequence down to 0.03\% and maintain a steady state temperature below 188\,mK at the temperature sensor.
We estimate a lower limit of the introduced heat load out of the calibration data of the cooling power for the dilution unit for this particular temperature to be 450\,$\mu$W.
Note that the experimental setup and the operation of the amplifier without running a pulse sequence already requires a cooling power of $\sim 39\,\mu$W.
\\
\begin{figure}[t!]
	\includegraphics[width=\textwidth]{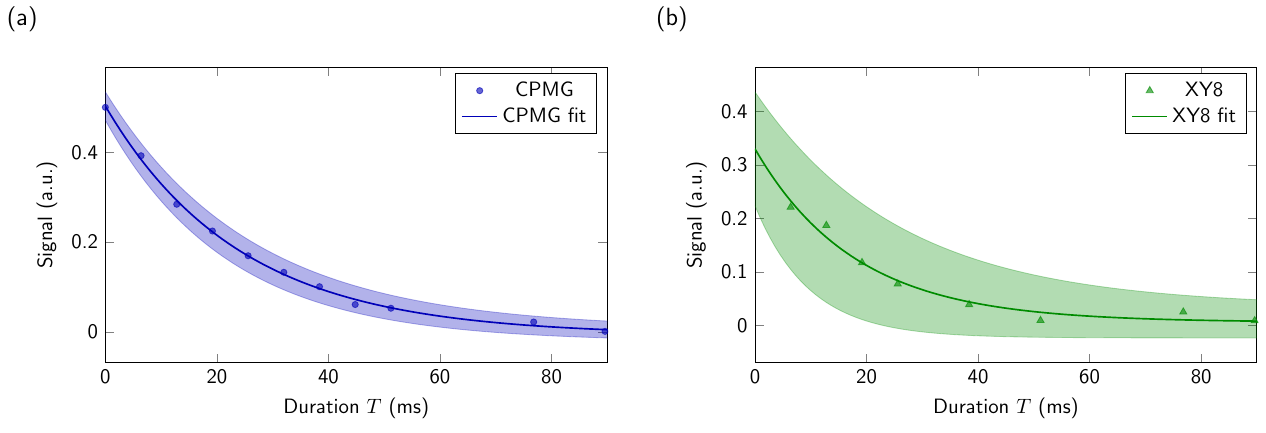}
	\caption{
		Differential memory measurements for CPMG (a) and XY8 (b).
		Data was retrieved by evaluation of first millisecond of the fluorescence trace of each laser pulse normalized by the maximum fluorescence of the spin flipped state.
		Fit shows well accordance to the data for both performed dynamical decoupling sequences and yields $T_{2,\text{CPMG}}=24.1\pm0.9,\text{ms}$ and $T_{2,\text{XY8}}=18\pm3\,\text{ms}$.
		Confidence intervals are depicted as shaded area.
		For the XY8 measurement the confidence interval appears broader, which we attribute to the less data points and the lower contrast.
		More details can be found in the text.
	}
	\label{Fig:CPMG_XY8_N_fit}
\end{figure}
Figure \ref{Fig:CPMG_XY8_N_fit} displays the measured decay curves for the CPMG and XY8 sequences, along with their corresponding fits and confidence intervals.
To ensure comparability among measurements of the different decoupling protocols, e.g. due to potential variations in photon counts caused by different excitation powers, a normalization of the measured signal was conducted.
For this purpose, a normalization sequence was implemented prior to every sequence, involving two 9 ms long laser pulses:
One pulse confirmed the counts in the dark state, while another, preceded by a $\pi$-pulse, determined the maximum count rate of the flipped state.
During the evaluation, only the accumulated counts of the first millisecond of each laser pulse were considered, as they carry essential information about the spin state (as shown in Fig. 1(d) of the main text).
These counts were then normalized by the counts of a full spin flip obtained from the normalization sequence.
Moreover, the conducted sequences were measured in an alternating manner, as depicted in Fig. 4(a) in the main text. 
To minimize statistical errors in the readout, each sequence was performed $10^4$ times.\\
As depicted in Fig. \ref{Fig:CPMG_XY8_N_fit}, the fit closely aligns with the data for both the CPMG and XY8 measurements, yielding respective results of $T_{2,\text{CPMG}}=24.1\pm0.9,\text{ms}$ and $T_{2,\text{XY8}}=18\pm3\,\text{ms}$.
However, the confidence interval appears comparatively larger for the XY8 sequence.
Possible reasons for this discrepancy could be the smaller number of measured points for XY8 compared to CPMG.
Moreover, the evaluated data for the XY8 measurements consistently exhibits a lower signal compared to the CPMG measurements.
We assume that this can be attributed to some contrast losses for the chosen $\tau$.
Fig. \ref{Fig:CPMG_XY8_Tau} shows some exemplary normalized differential signal from a CPMG-8 and a XY8 measurement, depicting this effect. 
For $\tau=100\,\mu$s (marked with a dashed vertical line) the signal shows for both sequences a dip, however this is more pronounced for the XY8 sequence.
Thus for the presentation of the data in the main text (Fig. 4(d)) we performed some re-normalization to account for this effect.\\  
It is important to emphasize that we achieve long coherence times and a strong agreement between the performed simulation and the measured data without further optimization of $\tau$.
We note that our simulations show that it is in principle possible to extend the coherence time even further, e.g., by even an order of magnitude, by reducing the inter-pulse waiting time, e.g., to 20-30 $\mu$s (see Eq. \eqref{Eq:OU_formula_full}), if a steady state temperature of 300\,mK could be maintained. In addition, the inter-pulse waiting time should be tailored to avoid resonances that would induce coupling to the $^{13}$C bath for the specific sequence, e.g., due to spurious harmonics \cite{Loretz2015PRX}.
However, these require a further optimization of the cooling process, a detailed investigation of the noise spectra of the spin bath, and most likely application of higher-order error-compensating DD sequences \cite{ajoy2011optimal,Genov2017PRL,ezzell2023dynamical}, which goes beyond the scope of this work.

\begin{figure}[t!]
	\includegraphics[height=0.2\textheight]{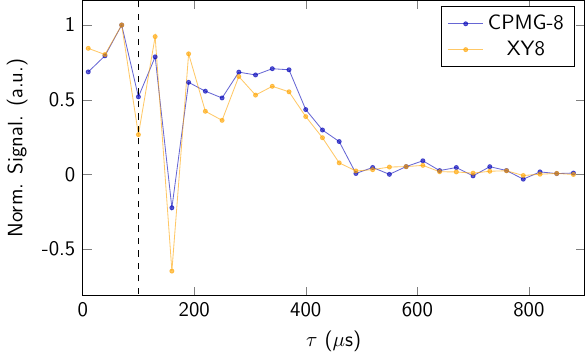}
	\caption{
		Decay curves for a differential measurement of CPMG-8 and XY8 for increasing interpulse duration $\tau$. 
		For $\tau=100\,\mu$s a dip for both dynamical decoupling sequences can be observed, most likely due to a nearby resonance and a resulting non-zero coupling to the $^{13}$C bath, which is more pronounced for the XY8 measurement \cite{Loretz2015PRX}. 
		This can have an impact on the quality of the measured signal displayed in Fig. \ref{Fig:CPMG_XY8_N_fit}.
		However, it is worth noting that we still achieve a strong agreement between the simulation and experimental results, even without optimizing $\tau$.
	}
	\label{Fig:CPMG_XY8_Tau}
\end{figure}

	
\end{document}